\RequirePackage{fix-cm}
\documentclass[smallextended]{svjour3}       % onecolumn (second format)
\smartqed  % flush right qed marks, e.g. at end of proof
\usepackage{graphicx}
\usepackage{natbib}
\usepackage{url}
\usepackage{rotating}
\usepackage{amsmath}
\usepackage{amssymb}
\journalname{Archived Article}
\begin{document}

\title{On the Privacy of Mental Health Apps \thanks{The work has been supported by the Cyber Security Cooperative Research Centre (CSCRC) Limited, whose activities are partially funded by the Australian Government’s Cooperative Research Centres Programme.}
}
\subtitle{An Empirical Investigation and Its Implications for Apps Development}


\author{Leonardo Horn Iwaya \and M. Ali Babar \and Awais Rashid \and Chamila Wijayarathna}


\institute{Leonardo Horn Iwaya \at
              1 Centre for Research on Engineering Software Technologies, the University of Adelaide, Adelaide, SA, Australia, 5005. \\
              2 Cyber Security Cooperative Research Centre (CSCRC), Australia. \\
              3 Privacy and Security (PriSec), Department of Mathematics and Computer Science, Karlstad University, Karlstad, Sweden. \\
              \email{leonardo.iwaya@kau.se}           %  \\
           \and
           M. Ali Babar \at
              1 Centre for Research on Engineering Software Technologies, the University of Adelaide, Adelaide, SA, Australia, 5005. \\
              2 Cyber Security Cooperative Research Centre (CSCRC), Australia.
          \and
          Awais Rashid \at
              4 Bristol Cyber Security Group, Department of Computer Science, University of Bristol, United Kingdom.
          \and
          Chamila Wijayarathna \at
              1 Centre for Research on Engineering Software Technologies, the University of Adelaide, Adelaide, SA, Australia, 5005. \\
              2 Cyber Security Cooperative Research Centre (CSCRC), Australia.
}

\date{Received: date / Accepted: date}

\maketitle

\begin{abstract}
An increasing number of mental health services are offered through mobile systems, a paradigm called mHealth. 
Although there is an unprecedented growth in the adoption of mHealth systems, partly due to the COVID-19 pandemic, concerns about data privacy risks due to security breaches are also increasing. 
Whilst some studies have analyzed mHealth apps from different angles, including security, there is relatively little evidence for data privacy issues that may exist in mHealth apps used for mental health services, whose recipients can be particularly vulnerable.
This paper reports an empirical study aimed at systematically identifying and understanding data privacy incorporated in mental health apps. 
We analyzed 27 top-ranked mental health apps from Google Play Store.
Our methodology enabled us to perform an in-depth privacy analysis of the apps, covering static and dynamic analysis, data sharing behaviour, server-side tests, privacy impact assessment requests, and privacy policy evaluation.
Furthermore, we mapped the findings to the LINDDUN threat taxonomy, describing how threats manifest on the studied apps.
The findings reveal important data privacy issues such as unnecessary permissions, insecure cryptography implementations, and leaks of personal data and credentials in logs and web requests.
There is also a high risk of user profiling as the apps' development do not provide foolproof mechanisms against linkability, detectability and identifiability.
Data sharing among third parties and advertisers in the current apps' ecosystem aggravates this situation.
Based on the empirical findings of this study, we provide recommendations to be considered by different stakeholders of mHealth apps in general and apps developers in particular.
We conclude that while developers ought to be more knowledgeable in considering and addressing privacy issues, users and health professionals can also play a role by demanding privacy-friendly apps.
\end{abstract}

\section{Introduction}
\label{sec:introduction}
The ongoing COVID-19 pandemic has dramatically increased the number of mental health support services provided using application developed for mobile devices. Such applications are called mental health apps, a subcategory of mobile health (mHealth) systems, such as chatbots (e.g., Wysa and Woebot), and text-a-therapist platforms (e.g., TalkSpace and BetterHelp), can be readily downloaded from apps stores, e.g., iOS or andriod, and used for seeking and/or providing help for mental health well-being \citep{echalliance,heilweil2020apps}.
Even before the COVID-19 pandemic, these apps make the provision of mental health services more accessible to the people in need, by lowering cost, eliminating traveling, saving time and reducing the fear of social stigma/embarrassment attached to psychological treatment \citep{bakker2016mental,price2014mhealth}.
Furthermore, mental health apps increase the availability of mental health services (``anywhere and anytime'') to users and provide additional functionalities such as real-time monitoring of users \citep{donker2013smartphones}. 
Research also shows that mental health apps improve users' autonomy and increase self-awareness and self-efficacy \citep{prentice2014review} leading to better health outcomes.

On the other hand, studies on the security of mHealth apps, in general, have shown that many apps are insecure, threatening the privacy of millions of users \citep{papageorgiou2018security}. Insecure apps can be the prime targets of cyber attackers since personal health information is of great value for cyber-criminals \citep{ibm2020cost}. There is also increasing evidence pointing to a widespread lack of security knowledge among mHealth developers, which is usually linked to different issues, such as insufficient security guidelines, tight budgets and deadlines, lack of security testing, and so on \citep{aljedaani2020empirical,aljedaani2021challenges}. 
App developers also heavily rely on a range of SDKs for analytics and advertising, exacerbating the risks of data linkage, detectability, and re-identification of users in such ecosystems \citep{solomos2019talon}.

The real or perceived security risks leading to data privacy compromises are particularly concerning for mental health apps because they deal with highly sensitive data, in contrast to other general mHealth apps, e.g., for fitness and wellness. The stigma around mental illnesses also increases the negative impacts on users in case of privacy violations. For instance, the mere link of users to a given app can reveal that they might be having some psychological problems (e.g., anxiety, depression, or other mental health conditions), which may make mental health apps users feel more vulnerable and fragile. 

The above-mentioned mHealth apps' data privacy concerns warrant evidence based inquiries for improved understanding and actionable measures as there is a paucity of empirical research on understanding the full range of privacy threats that manifest in mental health apps; the existing research has only focused on privacy policy analysis \citep{o2019reviewing,powell2018complexity,robillard2019availability,rosenfeld2017dementia}, or third party data sharing \citep{huckvale2019assessment}. Hence, it is important to systematically identify and understand the data privacy problems that may exist in mHealth apps as such a body of knowledge can better inform the stakeholders in general and apps developers in particular.

This study was stimulated by one research question: \emph{What is the current privacy status of top-ranked mental health apps?}
Here, we adopt a broad definition of privacy that encompasses security and data protection and with emphasis on the negative privacy impacts on data subjects.

The methodology for this investigation relied on a range of penetration testing tools and methods for systematic analysis of privacy policies and regulatory compliance artefacts.
We selected a sample of 27 top-ranked mental health apps from the Google Play Store that collected, stored and transmitted sensitive personal health information of users.
We subjected the apps to static and dynamic security analysis and privacy analysis by employing various tools such as MobSF, Drozer, Qualys SSL Labs, WebFX, CLAUDETTE and PrivacyCheck. Furthermore, we documented the privacy issues that we identified for each app by mapping them to the well-known LINDDUN privacy threat categories \citep{deng2011privacy} (i.e., \textbf{L}inkability, \textbf{I}dentifiability, \textbf{N}on-repudiation, \textbf{D}etectability, \textbf{D}isclosure of information, \textbf{U}nawareness and \textbf{N}on-compliance.

This study's findings reveal alarming data privacy problems in the mHealth apps used by millions of users, who are likely to expect data privacy protection built in such apps. Our study's main findings include:
\begin{itemize}
    \item Most apps pose linkability, identifiability, and detectability threats. This is a risk as some 3rd-parties can link, re-identify and detect the users' actions and data. Unawareness is also related to such threats, given that apps do not explain (e.g., in the privacy policy) the risks posed by targeted advertising on people experiencing mental problems and the risk of re-identification and disclosure of mental health conditions (e.g., anxiety, depression).
    \item Only 3/27 app developers responded to our query regarding PIAs, mentioning that they had performed a PIA on their app, and only two of them had made the PIA reports public. That suggests a high non-compliance rate since mhealth apps tend to pose high-risk to the right and freedoms of users.
    \item 24/27 app privacy policies were identified to require at least college-level education to understand them. The remaining 3/27 apps needed 10th--12th-grade level education to understand them. Such findings also suggest further problems with regards to non-compliance, leading to data subject's unawareness about the nature of the data processing activities in mental health apps, data controllers, and service providers.
    \item Static analysis reports show that 20/27 apps are at critical security risk, and 4/27 apps are at high security risk. Most issues are revealed through a simple static analysis, such as the use of weak cryptography. Dynamic analysis also shows that some apps transmit and log personal data in plain-text. Four apps can leak such sensitive data to 3rd-parties, exacerbating risks of (re-)identification and information disclosure.
\end{itemize}
We have also synthesised the main findings and mapped them according to the LINDDUN privacy threat taxonomy \citep{deng2011privacy}. The findings highlight the prevalence of data privacy problems among the top-ranked mHealth apps. It is clear that companies and software developers should pay more attention to privacy protection mechanisms while developing mHealth apps.
At the same time, users and mental health practitioners should demand for (at least) compliance with privacy standards and regulations.
Based on the findings, we offer some recommendations for mhealth apps development companies, apps developers, and other stakeholders.

\section{Background}

\subsection{Privacy (and Security)}
Until quite recently, the term privacy was treated under the umbrella of security. However, this situation has changed with data privacy gaining significance and prominence of it's own. It is essential to clarify the difference between privacy and security for the research reported in this paper.
In this study, we are mainly interested in data privacy that can be compromised as a result of security breach.
The concept of privacy comprises several aspects such as informed consent, transparency, and compliance, that are not necessarily connected to security.
Whilst privacy is protected through security measures, privacy cannot be satisfied solely on the basis of managing security \citep{nist8062}.
For such reasons, we regard security as part of a broad conceptualisation of privacy, which includes protecting personal data.
As a consequence, the study design reflects this contrast between privacy and security.
That is, apart from traditional security testing, this study also evaluates the apps' privacy policies, makes requests for privacy impact assessments, and gathers the developers' feedback on raised issues.

\subsection{The Ecosystem of Mental Health Apps}
\label{sec:mhealth-apps}
\begin{figure}[h]
  \centering
  \includegraphics[width=1.0\linewidth]{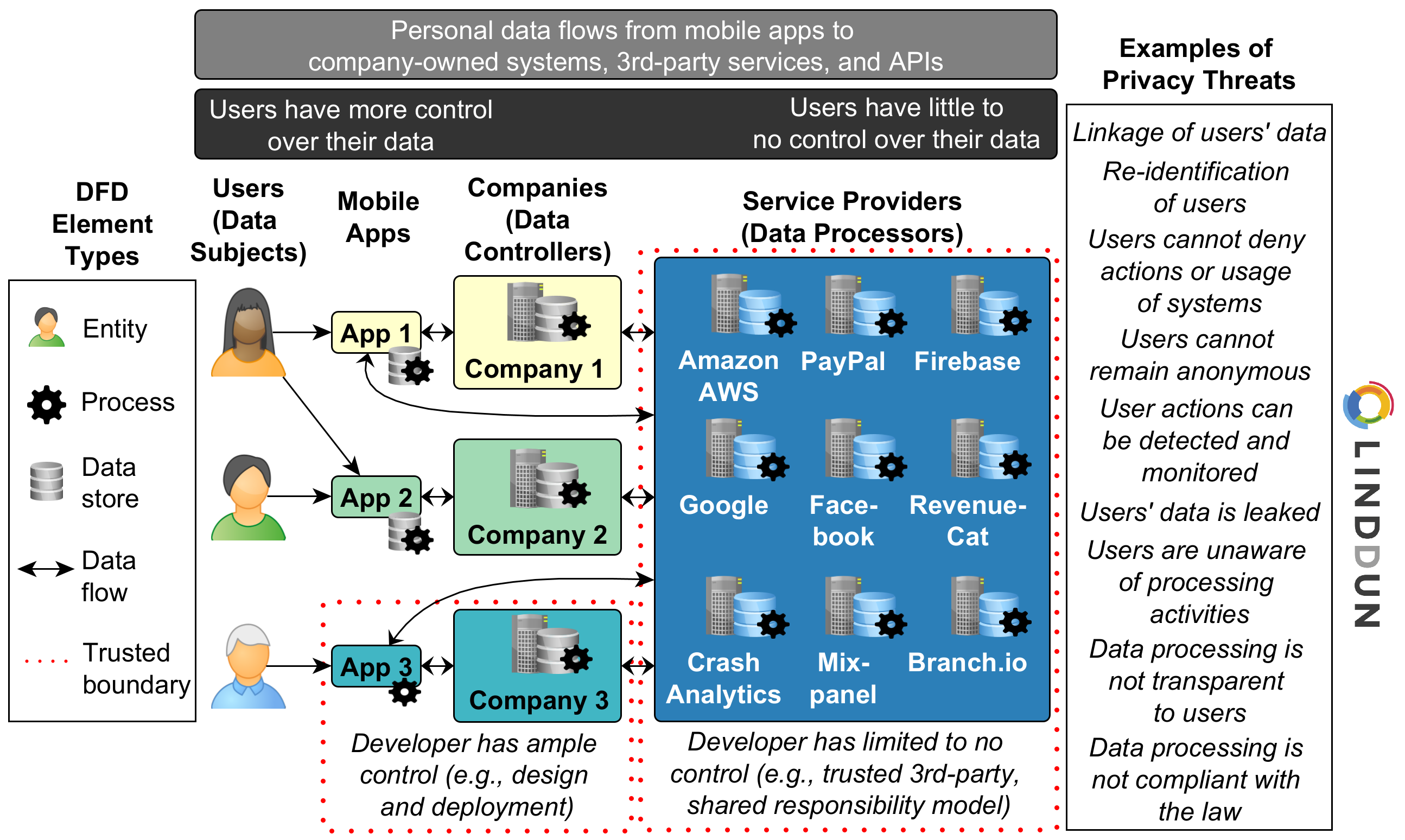}
  \caption{Simplified Data Flow Diagram (DFD) for the apps' ecosystem with an overview of the data subjects, data controllers, data processors, and privacy threats to consider.}
  \label{fig:apps-ecosystem}
\end{figure}

Today's information systems are built upon a wide range of services involving multiple stakeholders.
Figure \ref{fig:apps-ecosystem} presents a simplified Data Flow Diagram (DFD) that can help a reader to identify the main actors in the mental health apps ecosystem for discussing the privacy issues. As shown in the Figure \ref{fig:apps-ecosystem}, users (i.e., data subjects) have their data collected by mHealth apps and transmitted to the companies (i.e., data controllers) as well as to the other service providers (i.e., data processors).
Privacy considerations should be made for every step of the DFD (i.e., a detailed DFD created by apps developers) in which personal data is processed, stored and transmitted.

First, as shown in Figure \ref{fig:apps-ecosystem}, the personal data flows from the app to a company-owned server.
Here developers have a greater control on the system's design so that the main concern is the protection of data at-rest, in-transit and in-use.
Developers can fully understand all aspects of the company-owned infrastructure (i.e., client and server sides). 
Thus, they can transparently communicate the nature of personal data collection and processing to its users. Data flows within this trusted boundary of the company-owned systems tend to be less problematic regarding privacy.
However, it is worth stressing that privacy goes beyond data protection, so other privacy aspects should be considered, such as unawareness and non-compliance threat categories.

Second, personal data flows to many 3rd-party service providers that support the collection and processing of the users' data.
Most companies rely (often entirely) on public cloud infrastructures (e.g., Amazon AWS, Google Cloud) to maintain their servers and databases, as well as use many APIs that provide services for the apps to function (e.g., CrashAnalytics, RevenueCat, PayPal, Firebase).
In such cases, developers have limited control over the system, and the processing activities are not fully transparent anymore. Developers have to trust service providers, and a shared responsibility model ensues.
Thus, the data flows going to service providers should be carefully scrutinized. This concern is particularly critical in mental health apps since the personal data is considered highly sensitive as previously mentioned.

Adding to the problem, companies often rely on advertising as a source of monetary income for their apps, and mental health apps are no exception in such business models.
Thus, a user's information provided for using an app may be distributed to the app developer(s), to 3rd-party sites used for functionality reasons, and to unidentified 3rd-party marketers and advertisers \citep{giota2014ethical}.
Whilst users and health professionals are expected to be aware of such risks, it is important that companies' development practices result in mHealth apps that are transparent that they operate under such a business model.
Users already have little control over their data that resides within the developers' systems, let alone the data shared with 3rd-parties, such as mobile advertising platforms and data brokers.

\subsection{The LINDDUN Threat Taxonomy}
\label{sec:linddun-defs}
LINDDUN is a well-known privacy threat modelling framework \citep{deng2011privacy}, recently included in the NIST Privacy Framework \citep{nist2022privacy}. Given the increasing popularity of LINDDUN framework for systematically analyzing privacy threats during software systems development, we decided to use LINDDUN to analyze and map the findings from our study. The LINDDUN privacy threat analysis methodology consists of three main steps:
(1) modelling the systems, using DFDs and describing all data; 
(2) eliciting privacy threats, iterating over the DFD elements to identify threats using a taxonomy; and, 
(3) managing the threats, finding suitable solutions to tackle the uncovered threats.

We are mainly interested in the LINDDUN threat taxonomy, which can be used as a standard reference for discussing about privacy threats:
\begin{itemize}
    \item Linkability: an adversary can link two items of interest (IOI) without knowing the identity of the data subject(s) involved (e.g., service providers are able to link data coming from different apps about the same data subject).
    \item Identifiability: an adversary can identify a data subject from a set of data subjects through an IOI (e.g., service providers can re-identity a user based on leaked data, metadata, and unique IDs).
    \item Non-repudiation: the data subject cannot deny a claim, such as having performed an action or sent a request (e.g., data and transactions stored by companies and service providers cannot be deleted, revealing the users' actions).
    \item Detectability: an adversary can distinguish whether an IOI about a data subject exists or not, regardless of being able to read the contents itself (e.g., attackers can detect that a user's device is communicating with mental health services).
    \item Disclosure of information: an adversary can learn the content of an IOI about a data subject (e.g., data is transmitted in plain-text).
    \item Unawareness: the data subject is unaware of the collection, processing, storage, or sharing activities (and corresponding purposes) of the data subject’s data (e.g., the companies' privacy policy is not easy to understand and transparent about the nature of data processing).
    \item Non-compliance: the processing, storage, or handling of personal data is not compliant with legislation, regulation, and policy (e.g., companies failed to perform a Privacy Impact Assessment for their systems).
\end{itemize}

\subsection{Related Work}
\label{sec:related-work}
The security and privacy aspects of mHealth apps have been investigated by several studies such as \citep{dehling2015exploring} and \citep{papageorgiou2018security}.
However, these studies primarily include wellness and fitness apps instead of apps with highly sensitive data such as those in the mental health area. As shown in Table \ref{tab:related-work}, we could identify only eight studies related to the security and privacy of mental health apps.

\begin{sidewaystable}
\caption{Comparison of the existing works on privacy and/or security for mental health apps.}
\begin{center}
\begin{tabular}{p{0.1\textwidth}cp{0.04\textwidth}p{0.075\textwidth}p{0.425\textwidth}p{0.2\textwidth}}
\hline
\textbf{Ref} & \textbf{Year} &\textbf{N. of Apps} & \textbf{Condition} & \textbf{Privacy \& Security Scope of Analysis} & \textbf{Limitations} \\
\hline
\citep{huang2017permissions} & 2017 & 274 & Anxiety & Apps permissions. & (i) Limited to anxiety apps. (ii) Analyzes only apps' permissions. \\ 
\citep{huckvale2019assessment} & 2019 & 36 & Depression and smoking cessation & Evaluating privacy policy content. Assessment of data transmission. & (i) Limited to depression and smoking cessation apps. (ii) Analyzes only the privacy policies and network traffic. \\ 
\citep{muchagata2019dementia} & 2019 & 18 & Dementia & GDPR compliance criteria (i.e., installation (data types and permissions); privacy policy; terms and conditions; request for consent; special categories of data (explicit consent); portability of personal data, and the right to be forgotten) & (i) Limited to dementia apps. (ii) Analyzes only the apps' permissions and performs a GDPR compliance check. \\ 
\citep{o2019reviewing} & 2019 & 116 & Depression & Privacy policy ``transparency score''. & (i) Limited to depression apps. (ii) Analyzes only the privacy policies. \\ 
\citep{parker2019private} & 2019 & 61 & Mental health & Critical content analysis of promotional materials (includes apps' permissions and privacy policies). & (i) Analyzes only the apps' permissions and privacy policies. \\ 
\citep{powell2018complexity} & 2019 & 70 & Diabetes and mental health & Multiple metrics were used to evaluate the complexity of the app privacy policies. & (i) Analyzes only the complexity of privacy policies. \\ 
\citep{robillard2019availability} & 2019 & 369 & Track and mood & Apps were assessed for availability of a privacy policy and terms of agreement and if available, these documents were evaluated for both content and readability. & (i) Analyzes only the privacy policies and terms of agreement. \\ 
\citep{rosenfeld2017dementia} & 2017 & 33 & Dementia & Evaluating privacy policy content. & (i) Limited to dementia apps. (ii) Analyzes only the privacy policies. \\
\hline
\multicolumn{2}{c}{\textbf{This Work}} & 27 & Mental health & Extensive privacy (and security) analysis: request of PIAs from developers; privacy policy readability analysis; automated analysis of unfair clauses and GDPR compliance; penetration testing with static and dynamic analysis of the apps; analysis of reverse engineered code and apps' generated data. & (i) Limited to top-ranked mental health apps. (ii) Privacy policies analyzed using AI-assisted tools. \\
\hline
\end{tabular}
\label{tab:related-work}
\end{center}
\end{sidewaystable}

However, the related work (see Table \ref{tab:related-work}) has a limited scope of analysis. Most researchers focus only on the apps' privacy policies \citep{o2019reviewing, powell2018complexity, robillard2019availability, rosenfeld2017dementia}.
Another work investigates only the apps' permissions \citep{huang2017permissions}, or the combination of apps' permissions and privacy policies \citep{parker2019private}.
Another study \citep{muchagata2019dementia} proposes a scope of analysis to check for GDPR compliance, i.e., assessing the types of collected data, apps' permissions, and evidence of consent management, data portability and data deletion features.
Such approaches mostly identify Unawareness and Non-compliance issues, missing the other categories of privacy threats. That means their results do not have the depth of penetration tests to support the presence of the concrete privacy threats.

One study has also examined the apps' network traffic and data transmissions, in addition to assessing the privacy policies \citep{huckvale2019assessment}.
Looking into the network traffic enabled the identification of data that is transmitted to 3rd parties, such as marketing and advertising services.
To some extent, this study may cover all LINDDUN threat categories, but it misses many branches in the LINDDUN threat trees.
For instance, logs and stored data are not inspected for data leaks and weak access control, nor reverse engineered code is reviewed for insecure coding.
These types of inspections are important in order to achieve breadth and depth of privacy analysis.

In this work, we employed an extensive assessment framework for the privacy analysis of mental health apps, detailed in Section \ref{sec:methodology}.
In brief, our privacy analysis work included a series of penetration tests, with static and dynamic analysis, inspecting apps' permissions, network traffic, identified servers, reverse-engineered code, databases and generated data, which had not been explored in the related work shown in Table \ref{tab:related-work}.
Furthermore, the proposed privacy analysis also involves communication with companies and software developers by requesting the PIAs of the apps and discussing findings through the responsible disclosure process.

\section{Methodology}
\label{sec:methodology}
\begin{figure}[h]
  \centering
  \includegraphics[width=1.0\linewidth]{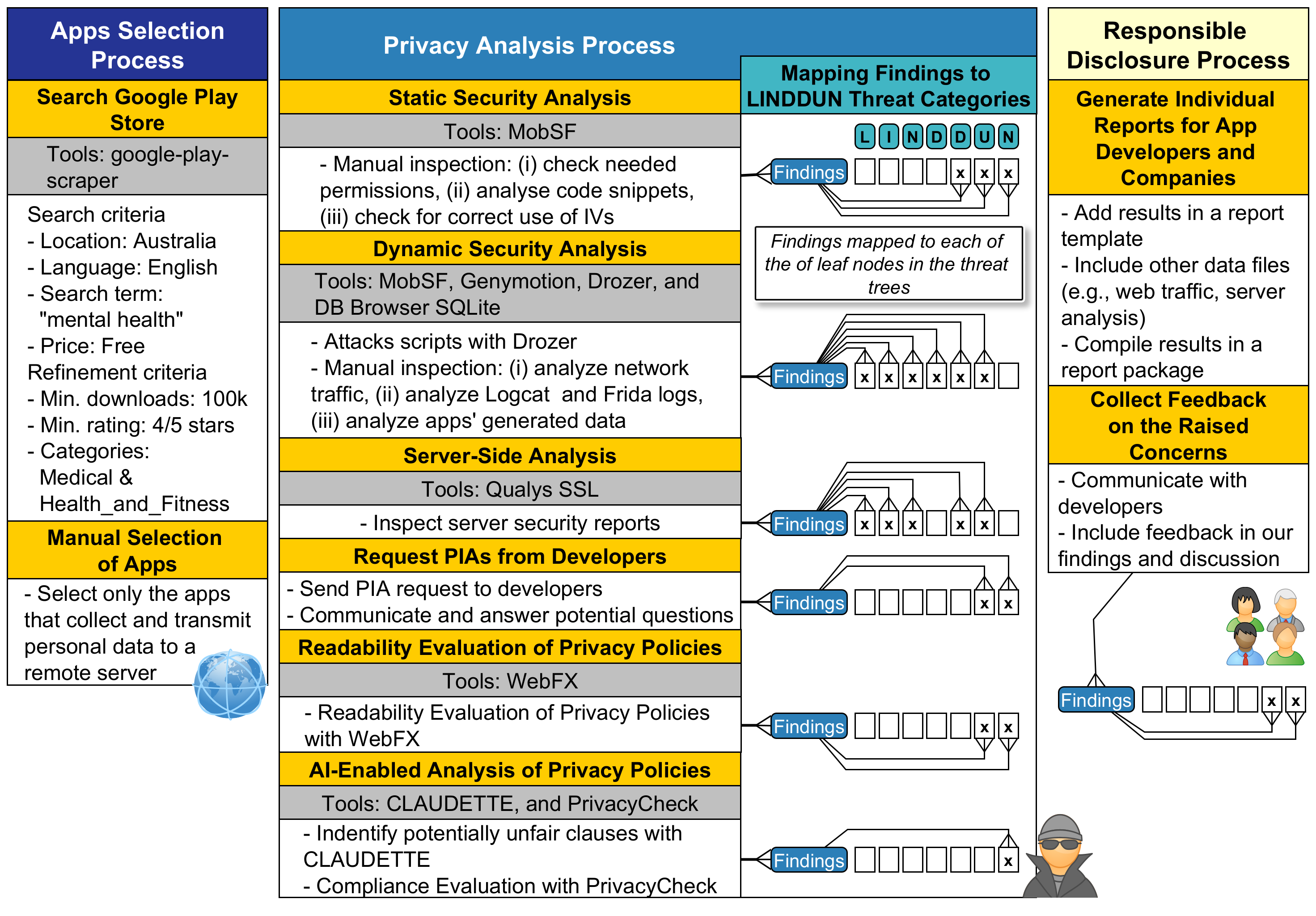}
  \caption{Methodology overview of the empirical investigation of privacy issues in mental health apps.}
  \label{fig:methodology-overview}
\end{figure}

This section presents the methodology used for the privacy assessment of the mental health apps in this study.
Figure \ref{fig:methodology-overview} gives an overview of all the steps followed for this study. 

\subsection{Apps Selection Process}
In this analysis, we used mobile applications developed for Android devices that are available to download in Google Play Store.
We performed the initial identification of the potential apps for this study using the \texttt{google-play-scraper} Node.js module\footnote{The \texttt{google-play-scraper} is a Node.js module to scrape application data from the Google Play store. Website:  \url{https://www.npmjs.com/package/google-play-scraper}}, essentially searching for free mental health apps in English (see Figure \ref{fig:methodology-overview}).

This search resulted in 250 apps as 250 is the default maximum set by Google Play Store. In order to select only the top-ranked apps, we introduced the following refinement criteria during the app selection process: apps should have at least 100K downloads, rating above 4 stars, and categorized as \texttt{MEDICAL} or \texttt{HEALTH\_AND\_FITNESS}. This refinement reduced our sample to 37 Android apps. 
We wanted to limit our analysis of the apps that require health and/or personal data as inputs in order to be functional and transmit users' data to a remote host. 
To identify these types of apps, we manually inspected the apps to figure out whether they store and transmit personal data of their users. 
This manual analysis included several tasks such as downloading the apps, reading their descriptions, creating and using dummy accounts to use the apps, manual entering information and checking apps' functionalities.
The analysis was stopped once sufficient evidence was gathered such as the app stored at least one personal data item and transmitted it.

This analysis identified nine apps that do not collect and transmit personal/health data of users and one app that would reveal sensitive information of other users if we performed analysis on that app. 
Therefore, we omitted these 10 apps from our analysis and selected the remaining 27 apps to perform the privacy-centered security analysis.

\subsection{Privacy Analysis Process}
\label{sec:methodology-privacy}
As shown in Figure \ref{fig:methodology-overview}, after filtering the 27 apps to perform the analysis, we performed static and dynamic security analysis to identify security vulnerabilities of the shortlisted mental health apps.
We also used Qualys SSL to evaluate all the servers identified during the dynamic security analysis.
Altogether, these three initial steps of the Privacy Analysis Process are mostly focused on the threats related to linkability, identifiability, non-repudiation, detectatbility and disclosure of information.
However, unawareness and non-compliance threats are also detected here but to a lesser extent, e.g., when analysing apps' permissions and manifest files.

In parallel, we also sent the PIA information requests for the studied apps to all developers/companies.
A readability analysis of the apps' privacy policies was conducted, and the apps were analyzed using AI-enabled tools to identify unfair clauses and points of non-compliance.
Hence, these remaining steps mainly targeted threat categories of unawareness and non-compliance.
All steps of the Privacy Analysis Process are detailed in the next sub-sections.

\subsubsection{Static Security Analysis}
\label{sec:methodology-static-analysis}
We performed the static analysis using an open-source analysis tool called MobSF, which is known for it's ease of use as it is fully automated \footnote{Mobile Security Framework (MobSF) is an automated, all-in-one mobile application (Android/iOS/Windows) pen-testing, malware analysis and security assessment framework capable of performing static and dynamic analysis. Website:  \url{https://mobsf.github.io/Mobile-Security-Framework-MobSF/}}. 
MobSF is widely used by security researchers for performing security analysis of mobile applications \citep{papageorgiou2018security}. 
Furthermore, previous research has shown MobSF's capability for identifying a wide range of Android security issues \citep{ranganath2020free}.

To perform the static analysis, we downloaded the APK file for each app and analyzed it using the MobSF static analyzer.
This analysis reveals various details about each app, e.g., including the apps' average Common Vulnerability Scoring System (CVSS) Score \citep{firstcvss}, trackers, certificates, android permissions, hard-coded secrets, and URLs etc.

One of the limitations of this type of analysis is that MobSF may report a considerable amount of false positives related to some vulnerabilities \citep{papageorgiou2018security}. 
Therefore, based on the initial results obtained by MobSF, we further performed the following checks to verify the issues reported by MobSF.

\begin{itemize}
    \item Manually evaluate whether the used ``dangerous'' permissions are required to serve the app's purpose.
    \item Manually analyse the code snippets that were reported to use insecure random number generators, insecure ciphers and insecure cipher modes.
    \item Manually checked the code snippets that used \texttt{IvParameterSpec} to test whether Initializing Vectors (IVs) have been correctly used.
\end{itemize}

\subsubsection{Dynamic Security Analysis}
\label{sec:methodology-dynamic-analysis}
As the next step, we performed the dynamic analysis of the apps. 
Dynamic Analysis is a black-box security testing methodology that analyzes an app by running it and performing potentially malicious operations on it. 
For performing dynamic analysis, we used Genymotion Android emulator\footnote{Genymotion is an emulator for Android devices. Website: \url{https://www.genymotion.com/}}, MobSF dynamic analyser and Drozer\footnote{Drozer offers a comprehensive security and attack framework for Android. Website: \url{https://labs.f-secure.com/tools/drozer/}}. 
Only 19 apps were subjected to the analysis in this step as the other eight apps were not compatible to run on the Android emulator with MobSF. 
We consider this as a limitation of the used methodology.

In the analysis process, we installed each of the studied apps into Genymotion emulator and manually performed various operations on each app while MobSF dynamic analyzer was listening to the performed operations. 
At the end of the analysis, MobSF provided us with a report that included the complete Logcat log, Dumpsys log, Frida API monitor log and HTTP/S traffic log for the whole period that we were interacting with each app. 
Furthermore, MobSF allowed us to download all the data created by each app that persisted in the device's storage.

In addition, when an app was running on the Android emulator, we used Drozer to perform malicious operations on the app to identify app's security vulnerabilities \citep{drozerguide}. 
We used various attack scripts such as checking for attack surfaces, SQL injection, and directory traversal vulnerabilities.

Thereafter, we performed a detailed analysis of the logs and apps' data files that were obtained from MobSF dynamic analysis. 
The HTTP/S traffic log provided the request and response information for each HTTP/S communication made during the process. 
We went through the log entries for each communication and investigated for any insecure channels that might have communicated users' sensitive data. Furthermore, we also checked for the 3rd-party servers that each app was sending users' personal and health data. 

In the next step, we analyzed Logcat logs and Frida API monitor logs generated during the dynamic analysis process to identify whether or not these logs reveal personal information of a user, reveal apps' behavior, usage, and activities, reveal tokens and credentials used by the app, and reveal the details of web traffic, parameters and Post values. 
Logcat logs generated in the device are accessible to other apps running on the device and logging sensitive information make such information accessible to those apps \citep{kotipalli2016hacking}. 
To identify these insecure log entries, we visually inspected all files and also performed a keyword search on log files with a list of keywords that included `username', `password', `API key', `key', `@gmail.com', etc.

As the final step of the dynamic analysis process, we analyzed the data generated by each app (i.e. files and databases) to see whether or not an app has insecurely stored a user's sensitive data. 
We categorized the data as encrypted or not encrypted, and used DB Browser SQLite\footnote{Database browser for SQLite databases. Website: \url{https://sqlitebrowser.org/}} to open and browse the data stored in the apps' databases folders. Similar to the log analysis step, we used visual inspection and keyword search to look for sensitive data that has been stored insecurely.

\subsubsection{Server-Side Analysis}
We also performed web server analysis on each domain with which the app communicated during the dynamic analysis. 
As part of this step, the relevant web servers' configurations were analyzed to assess the security levels of the HTTPS data transmissions.
To perform this analysis, we used Qualys SSL Labs\footnote{Qualys SSL is a free online service to perform a deep analysis of the configuration of any SSL web server on the public Internet. Website: \url{https://www.ssllabs.com/ssltest/}} tool, which is a free online service that enables the remote testing of web server's security against a number of well-known vulnerabilities, such as Heartbleed \citep{durumeric2014matter} and Drown \citep{aviram2016drown}. 
The analysis provided an overall rating for the web server's security (A+, A, B, C, D, E, F) as well as a score and weaknesses for certificate, protocol support, key exchange and cipher strength aspects.

\subsubsection{Request Privacy Impact Assessment}
Privacy Impact Assessment (PIA), also known as Data Protection Impact Assessment, is an important component of an app's accountability that comes under GDPR \citep{icogdpr}. 
Most information privacy regulations, such as the GDPR and the Australian Privacy Act, encourage the publication of PIA reports as it demonstrates to stakeholders and the community that the project has undergone critical privacy analysis, potentially reducing community concerns about privacy \citep{gdpreudpia,oaicpia}. 
Therefore, as a part of our privacy analysis, we evaluated whether or not the developers of the studied apps had performed PIA on their respective apps and made the findings public. 
We contacted the companies and/or developers of the studied apps based on the contact details available on Google Play Store and requested them to send the details of the public reports of their PIAs. 

\subsubsection{Readability Evaluation of Privacy Policies}
Privacy policies are responsible for communicating how an app gathers, uses, discloses, and manages the personal information of the app users \citep{zaeem2020effect}. 
Previous research has evaluated privacy policies of different types of apps and reported that privacy policies are often too complex and difficult for users to read and understand \citep{o2019reviewing,powell2018complexity}. 
We were interested in evaluating the readability of privacy policies of the mental health app as these apps are often used by users who are already psychologically and cognitively challenged \citep{marvel2004cognitive}.

Therefore, as a part of the privacy analysis step, we evaluated the readability of the apps' privacy policies. We used WebFX\footnote{The WebFX Readability Test Tool provides a way to test the readability of any textual content. Website: \url{https://www.webfx.com/tools/read-able}} free online tool for this. This tool provides various readability scores (e.g., Flesch-Kincaid, Gunning Fog, SMOG), as well as a number of metrics about the privacy policies (e.g., number of words, sentences, complex words).

\subsubsection{AI-Enabled Analysis of Privacy Policies}
\label{sec:methodology-privacy-compliance}
We performed the final component of the privacy policy analysis using two AI-enabled tools, which are CLAUDETTE \citep{lippi2019claudette} and PrivacyCheck \citep{zaeem2020effect,zaeem2018privacycheck}. 
First, we used CLAUDETTE to identify the potentially unfair clauses in apps' privacy policies, e.g., jurisdiction disputes, choice of law, unilateral termination or change.
In addition, we used PrivacyCheck, which is an automated tool provided as a Chrome browser plugin. 
It evaluates the privacy policy of an app with respect to 20 points criteria where 10 questions are related to users' control over their privacy and 10 questions are related to GDPR.

\subsection{Responsible Disclosure Process}
After completing the Privacy Analysis Process of the selected apps, we prepared the reports on the results for each app.
We emailed the evaluation reports to the companies and/or developers of the apps based on the contact details available on Google Play Store and asked them to respond within 30 days whether or not they had fixed the identified security and privacy issues.
We gathered the information about how they responded to our report and whether or not they improved their apps based on our findings.

\subsection{Mapping Findings to LINDDUN}
As the last step, a detailed mapping exercise was performed.
Essentially, throughout the privacy analysis, a list of privacy issues was compiled for each app. 
Then, we followed the knowledge support provided by the LINDDUN methodology for cross-checking every single issue in the list with respect to the entire threat taxonomy (threat-by-threat) to check for correspondence. 
Finally, if one of the threats is relevant to a given issue, this threat is mapped and included in the mapping table (readers are referred to the LINDDUN's threat tree catalog (v2.0) \citep{wuyts2014linddun-catalog} for consultation).

\begin{figure}[h]
  \centering
  \includegraphics[width=1.0\linewidth]{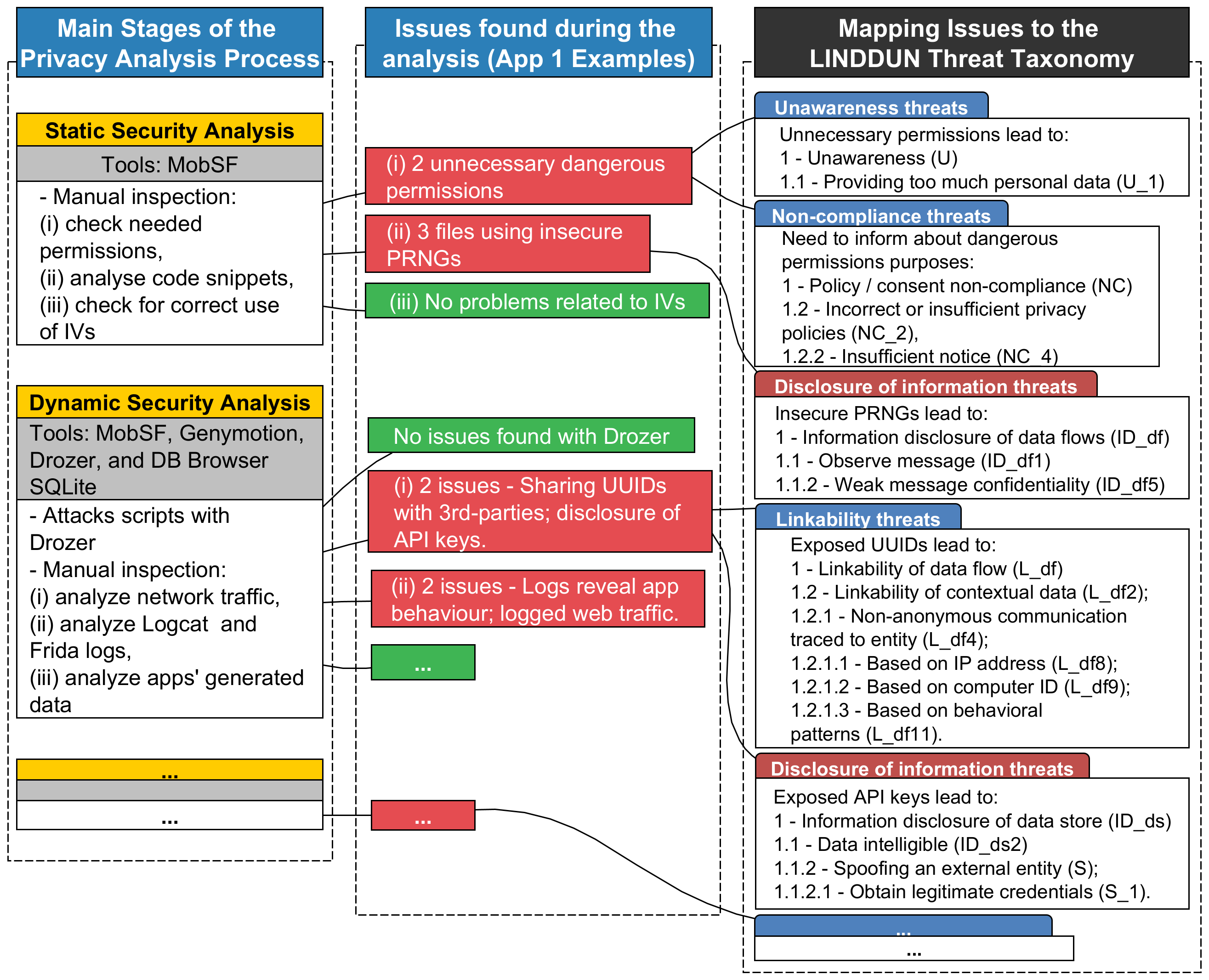}
  \caption{Example of mapping process for App 1: associating issues found during the analysis to the LINDDUN taxonomy threats.}
  \label{fig:mapping-example}
\end{figure}

An illustrative example is provided in Figure \ref{fig:mapping-example}.
Every step of the Privacy Analysis Process allows identifying a number of issues.
For instance, during the Security Static Analysis of App 1, two dangerous permissions were identified, and three files in the reversed engineered code used insecure PRNGs.
One dangerous permission is the \texttt{android.permission.READ\_PROFILE}, which allows the application to read the user's profile data.
This permission does not seem necessary at installation time nor for the app to function for its specified purposes, thus it was marked as an issue.
Having such dangerous permission results in providing too much data (i.e., Unawareness threat).
Also, it relates to insufficient notice to users (i.e., Non-compliance threat tree) since the privacy policy could have better explained the need for this dangerous permission.
Similarly, the issues regarding the use of insecure PRNGs may lead to insecure security implementations, and thus, weak message confidentiality (i.e., Disclosure of Information threat).
These overarching findings are presented in Section \ref{sec:results} along with the results.

\section{Results}
\label{sec:results}

\subsection{Selected Mental Health Apps}
\label{sec:selected-apps}
The final sample consists of 27 Android apps that provide functionalities related to mental health services. 
Twenty-one of the selected apps were from the `Health \& Fitness' genre; the remaining six apps were from the `Medical' genre.
Table \ref{tab:app-downloads} provides a summary of the sample of apps used in this study. The selected apps originate from 11 different countries from four continents.
To keep the apps de-identified, we have not included the exact details about the apps' origin countries.

Table \ref{tab:app-themes} provides the results of a tagging exercise performed by the researchers for all the selected apps.
Each app was tagged with several tags representing their scope, which allowed us to group them in themes.
Notice that an apps may fall into one or more themes.
As shown in Table \ref{tab:app-themes} Anxiety, Stress and Depression are the most common tags among the selected apps.
This tagging exercise provides an overview of the apps' themes while keeping the apps de-identified.

\begin{table}[htbp]
\caption{Apps with their respective number of downloads.}
\begin{center}
\begin{tabular}{cc}
\hline
\textbf{N. of Apps}&\textbf{N. of Downloads} \\
\hline
12 & 100,000 - 500,000 \\
3 & 500,000 - 1,000,000 \\
9 & 1,000,000 - 5,000,000 \\
1 & 5,000,000 - 10,000,000 \\
2 & 10,000,000+ \\
\hline
\end{tabular}
\label{tab:app-downloads}
\end{center}
\end{table}

\begin{table}[htbp]
\caption{Themes of Analysed apps.}
\begin{center}
\begin{tabular}{cp{0.8\linewidth}}
\hline
\textbf{N. of Apps} & \textbf{Tags}\\ \hline
22 & Anxiety \\ 
19 & Stress and burnout \\ 
13 & Depression \\ 
13 & Sleep and insomnia \\ 
13 & Journal, diary and daily-planner \\ 
12 & Mood and habit tracker \\ 
10 & Disorders, addiction, bipolar, anger, phobia, eating disorder, negative emotions, mood disorder, self-harm, PTSD, OCD, and ADHD \\ 
8 & Meditation \\ 
8 & Panic attack \\ 
8 & Online therapy, online doctor and couples therapy \\ 
5 & Chatbot \\ 
5 & Other, e.g., peer-support, pregnancy, pain management, bullying \\ 
4 & Self-esteem \\ 
3 & Mental health assessment, diagnosis and check symptoms \\ \hline
\end{tabular}
\label{tab:app-themes}
\end{center}
\end{table}

Of the 27 top-ranked mental health apps selected, most address the conditions of anxiety, stress, burnout and depression. 
Also, over a third of them address various other mental health conditions, e.g., addictions, bipolar, self-harm, PTSD and OCD. 
For these reasons, we argue that these apps' processing operations ought to be considered ``high-risk'' to the rights and freedoms of their users.

\subsection{Summary of Results According to LINDDUN}
\label{sec:mapping-summary}
This section summarises the mapping between the identified issues for a given app and the LINDDUN threat categories.
As shown in Table \ref{tab:map-to-linddun}, considering App 1, three of the found issues were mapped to one or more of the Linkability threats.
In what follows, we structure the results section around the seven LINDDUN threat categories, covering the main threats that manifest in the studied apps.
Evidence gathered during the Privacy Analysis Process, such as the results of tools (e.g., MobSF, Qualys SSL, CLAUDETTE) and manual analysis of network traffic and logs, are used as examples of how the threats appear.

\begin{table}[htbp]
\caption{Mapping summary, showing the number of times that one of the apps' issues was mapped to a threat category. Note: (*) means that the dynamic analysis could not be performed for the app.}
\begin{center}
\begin{tabular}{ccccccccc}
\hline
\textbf{App Code} & \textbf{L} & \textbf{I} & \textbf{N} & \textbf{D} & \textbf{D} & \textbf{U} & \textbf{N} & \textbf{Total} \\
\hline

App 1 & 3 & 3 & 3 & 3 & 5 & 5 & 7 & 29 \\
App 2 & 2 & 2 & 2 & 2 & 6 & 4 & 5 & 23 \\ 
App 3 & 2 & 2 & 2 & 2 & 7 & 4 & 6 & 25 \\ 
App 4* & - & - & - & - & 4 & 4 & 6 & 14 \\
App 5 & 2 & 2 & 2 & 2 & 7 & 4 & 5 & 24 \\ 
App 6* & - & - & - & - & 2 & 4 & 5 & 11 \\ 
App 7 & 3 & 3 & 3 & 3 & 6 & 5 & 7 & 30 \\
App 8* & - & - & - & - & 3 & 4 & 6 & 13 \\ 
App 9 & 2 & 2 & 2 & 2 & 6 & 4 & 6 & 24 \\ 
App 10 & 2 & 2 & 2 & 2 & 6 & 4 & 6 & 24 \\ 
App 11 & 3 & 3 & 3 & 3 & 7 & 4 & 5 & 28 \\ 
App 12 & 3 & 3 & 3 & 3 & 6 & 5 & 7 & 30 \\ 
App 13 & 3 & 3 & 3 & 3 & 6 & 5 & 7 & 30 \\
App 14 & 3 & 3 & 3 & 2 & 5 & 5 & 7 & 28 \\ 
App 15 & 3 & 3 & 3 & 3 & 6 & 5 & 7 & 30 \\ 
App 16 & 3 & 3 & 3 & 3 & 7 & 5 & 7 & 31 \\ 
App 17 & 3 & 3 & 3 & 3 & 6 & 5 & 7 & 30 \\ 
App 18 & 3 & 3 & 3 & 3 & 7 & 5 & 6 & 30 \\ 
App 19* & - & - & - & - & 2 & 4 & 6 & 12 \\
App 20 & 2 & 2 & 2 & 2 & 4 & 3 & 4 & 19 \\ 
App 21* & - & - & - & - & 1 & 4 & 6 & 11 \\ 
App 22* & - & - & - & - & 1 & 4 & 6 & 11 \\ 
App 23 & 3 & 3 & 3 & 3 & 7 & 5 & 7 & 31 \\ 
App 24* & - & - & - & - & 1 & 4 & 6 & 11 \\ 
App 25* & - & - & - & - & 2 & 4 & 6 & 12 \\
App 26 & 2 & 2 & 2 & 2 & 4 & 3 & 4 & 19 \\ 
App 27 & 2 & 2 & 2 & 2 & 5 & 4 & 6 & 23 \\
\hline
\textbf{Avgs.:} & 1.8 & 1.8 & 1.8 & 1.8 & 4.8 & 4.3 & 6.0 & 22.3  \\
\hline
\end{tabular}
\label{tab:map-to-linddun}
\end{center}
\end{table}

\subsubsection{Linkability Threats}
\label{sec:linkability-threats}
LINDDUN borrows most of its terminology definitions from the work of \citep{pfitzmann2010terminology}, including the definition for linkability.
Linkability is the ability to sufficiently distinguish whether two IOI are linked or not, even without knowing the actual identity of the subject of the linkable IOI.
Typical examples are anonymous letters written by the same person, web page visited by the same user, entries in two databases related to the same person, people related by a friendship link, etc.

Such linkability threats are revealed during the dynamic analysis of an apps when network traffic was manually inspected.
This is done using tools such as MobSF and Genymotion to emulate apps and capture their network traffic, logs, and generated data.

\begin{figure}[h]
  \centering
  \includegraphics[width=0.7\linewidth]{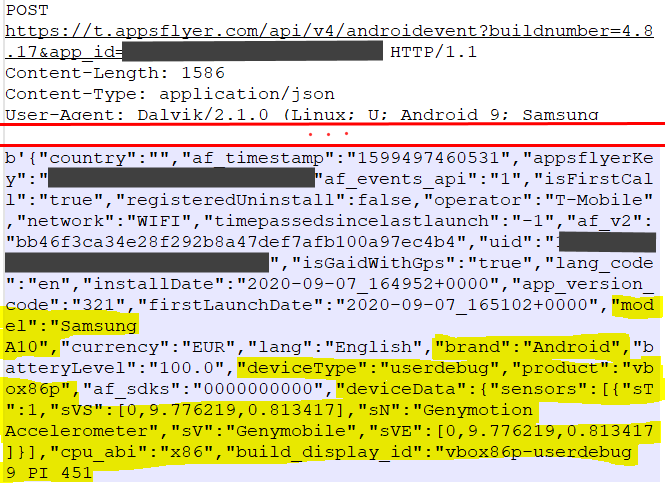}
  \caption{Example of 3rd-party receiving user's device information.}
  \label{fig:battery-3rd-party}
\end{figure}

The most prevalent type of threat refers to the linkability of contextual data (L\_df2) concerning data flows.
Contextual data becomes linkable when non-anonymous communication (L\_df4) is used, which is the reality for all the selected apps.
Hence, the data flow can be linked based on IP address, device IDs, sessions IDs, or even communication patterns (e.g., frequency, location, browser settings).
An example of an app leaking such data to 3rd-parties, such as users' activities in the app and the device configuration, is shown in Figure \ref{fig:battery-3rd-party}.

\begin{figure}[h]
  \centering
  \includegraphics[width=0.7\linewidth]{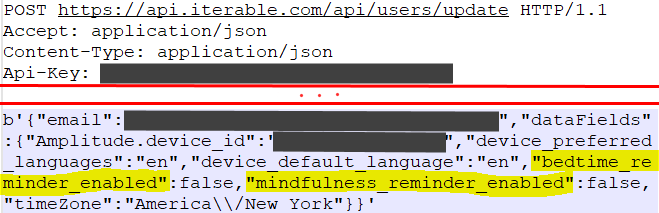}
  \caption{Example of 3rd-party receiving user's activities information.}
  \label{fig:activities-3rd-party}
\end{figure}

Such linkability threats manifested in all the 18 apps that went through the dynamic analysis.
User behaviour can be easily extracted from web traffic logs (i.e., it is easy to perform profiling of mental health apps’ users), even if one cannot re-identify a subject (see Figure \ref{fig:activities-3rd-party}).
Most apps also attempt to pseudo-anonymize users through anonymous IDs or hashed advertisement IDs, but these IDs can still be used to link data among various 3rd-parties.
In particular circumstances, two apps exacerbate linkability threats by generating a perplexing number of HTTP(S) requests in a short period (i.e., App 23 did 507 and App 15 made 1124 requests).
The more data is available, the worst it is in terms of linkability. More data points are linked over longer periods of time; and it's also harder to hide the links between two or more items of interest (e.g., actions, identifiers).

\subsubsection{Identifiability Threats}
\label{sec:identifiability-threats}
Identifiability of a subject from an attacker’s perspective means that an attacker can sufficiently identify a subject within a set of subjects \citep{pfitzmann2010terminology}.
Examples are identifying the reader of a web page, the sender of an email, the person to whom an entry in a database relates, etc.
It is worth mentioning that likability threats increase the risks of re-identification. 
The more information is linked, the higher the chance the combined data are identifiable (i.e., the more attributes are known, the smaller the anonymity set).

\begin{figure}[h]
  \centering
  \includegraphics[width=0.7\linewidth]{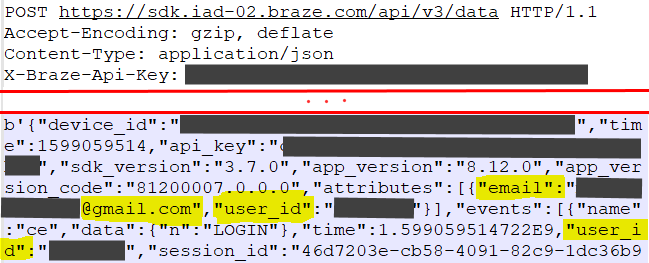}
  \caption{Example of 3rd-party receiving user's email information.}
  \label{fig:email-3rd-party}
\end{figure}

Identifiability threats are also revealed through the dynamic analysis when inspecting network traffic, logged and stored data, using tools such as MobSF, Genymotion, Logcat dumps, and DB Browser SQLite.
Here we are particularly interested in data flows that go to 3rd-parties or that may be accessible by attackers (i.e., situations in which users typically assume that they are anonymous).
Identifiability of log-in used (I\_e1) and contextual data (I\_df2) were the most common types of threat found in the 18 apps that went through dynamic analysis.
In such cases, users can be re-identified by leaked pseudo-identifiers, such as usernames and email addresses, as shown in Figure \ref{fig:email-3rd-party}.

Identifiability may also manifest due to weak access control to stored data (I\_ds1).
These situations were observed when apps' leak personal information in the system logs (accessible by all apps), or store data in plain text, using databases or external storage.
However, attackers would need physical access to the device to exploit such threats, and in such cases, it is likely that they already know the victim's identity.
Such types of threats are nonetheless discussed under the threat category of Disclosure of Information in Section \ref{sec:disclosure-info-threats}.

\subsubsection{Non-repudiation Threats}
\label{sec:non-repudiation-threats}
Non-repudiation refers to not being able to deny a claim or action.
Therefore, an attacker can prove that a user knows, has done, or said something, such as using mental health apps and services.
Here again, we are particularly interested in non-repudiation threats involving 3rd-party systems.

\begin{figure}[h]
  \centering
  \includegraphics[width=0.7\linewidth]{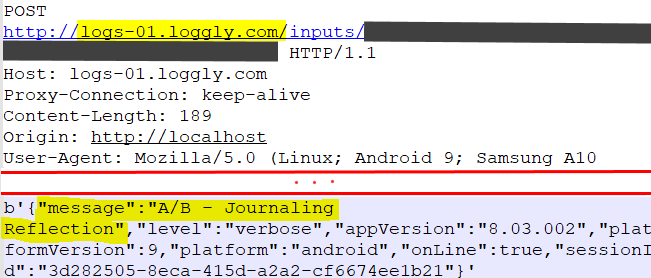}
  \caption{Example of 3rd-party logging service used to record apps' activities, sending data over HTTP.}
  \label{fig:pii-http-logs}
\end{figure}

Such threats are also identified during the dynamic analysis. We observed non-repudiation threats related to the disclosure of decrypted logs of network connections (NR\_df7), and when a person wanting deniability cannot edit a database (NR\_ds3).
The analyzed apps communicate with several 3rd-parties, e.g., for marketing and advertising, cloud service provisioning, and payments services. 
This makes it impossible for users to determine to what extent their communication and data are collected, used, and stored.
A rather worrying example is the logging of user actions in an app by a 3rd-party logging service using the insecure HTTP protocol, as shown in Figure \ref{fig:pii-http-logs}.

Table \ref{tab:web-servers} shows the number of servers that the apps communicated with during the analysis. 
On average, an app communicated with 11.9 servers ($std = 13.8$), with a minimum of 1 and a maximum of 64 communicating servers.
Most of these servers are 3rd-party service providers. On average, 81.7\% ($std = 18.3$) of the servers that each app communicated were owned by 3rd-parties. 
Such intense use of service providers increases the risks of non-repudiation. In addition, if the data that is shared is identifiable, it will be harder to repudiate.

\begin{table}[htbp]
\caption{Web servers communicated during the dynamic analysis.}
\begin{center}
\begin{tabular}{cc}
\hline
\textbf{N. of Apps} & \textbf{N. of Web Servers} \\
\hline
6 & 1-5 \\
5 & 6-10 \\
6 & 11-20 \\
2 & $>$20 \\
\hline
\end{tabular}
\label{tab:web-servers}
\end{center}
\end{table}

Table \ref{tab:3rd-party-domains} presents a list of the 3rd-party domains most commonly observed in the performed analysis. 
App developers use such common 3rd-parties for marketing (e.g., Mixpanel, RevenueCat, Branch.io, Amplitude, Facebook), cloud service provisioning (Firebase, CrashAnalytics, Bugsnag), and payment services (e.g., Stripe and PayPal). Software developers and users often have little to no control over the data after sharing it with service providers.

\begin{table}[htbp]
\caption{Most common 3rd-party domains.}
\begin{center}
\begin{tabular}{cl}
\hline
\textbf{N. of Apps} & \textbf{Domain} \\
\hline
18 & \texttt{google.com} \\
15 & \texttt{googleapis.com} \\
12 & \texttt{crashlytics.com} \\
9 & \texttt{branch.io} \\
8 & \texttt{facebook.com} \\
8 & \texttt{gstatic.com} \\
7 & \texttt{mixpanel.com} \\
7 & \texttt{youtube.com} \\
6 & \texttt{app-measurement.com} \\
\hline
\end{tabular}
\label{tab:3rd-party-domains}
\end{center}
\end{table}

\subsubsection{Detectability Threats}
\label{sec:detectability-threats}
Detectability refers to being able to sufficiently distinguish whether or not an IOI exists \citep{pfitzmann2010terminology}, even if the actual content is not known (i.e., no information is disclosed).
Based on the detection of whether or not an IOI exists, one can infer or deduce certain information.
For instance, by knowing that a user has a profile in a specific mental health service, you can deduce that they might be seeking psychological support or facing specific mental health conditions. Achieving undetectability in mobile and web applications is inherently complex, given that client-server communication is usually easily detectable.

All apps that generate network traffic present detectability threats.
Threats are observed during the dynamic analysis, such as no or weak covert channel (D\_df2), since data flows can be examined (D\_df7) and the timing of the requests is visible (D\_df13).
The data stored by the apps is also detectable due to the weak access control to the data file system or database (D\_df1).
Software developers cannot easily address such threats, considering that existing apps would have to provide relatively advanced privacy controls, such as using covert channels and anonymous communication.
The reliance on various 3rd-party service providers makes it even more challenging.

\subsubsection{Disclosure of Information Threats}
\label{sec:disclosure-info-threats}
Information disclosure refers to the unwanted and unauthorised revelation of information. 
For data flows, the channel is insufficiently protected (e.g., un-encrypted), and the message is not kept confidential.
Similarly, the information is protected with weak access control mechanisms or kept in plain text for data stored.

Threats on disclosure of information were observed in the static, dynamic, and server-side analyses, using MobSF to reverse engineer code and inspect data flows and server configuration.
Based on MobSF's static analysis, 74\% ($n = 20$) of the apps scored as Critical Risk and 15\% ($n = 4$) as High Risk in the App Security Score.
From the beginning of the analysis, such negative scores suggested that many apps would have problems in terms of permissions, code vulnerabilities, trackers, etc.

\begin{figure}[h]
  \centering
  \includegraphics[width=0.7\linewidth]{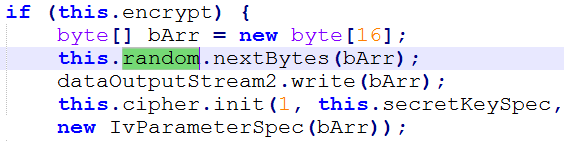}
  \caption{Example of insecure random (i.e., \texttt{java.util.Random}) used to generate IVs.}
  \label{fig:insecure-random-iv}
\end{figure}

\begin{figure}[h]
  \centering
  \includegraphics[width=0.8\linewidth]{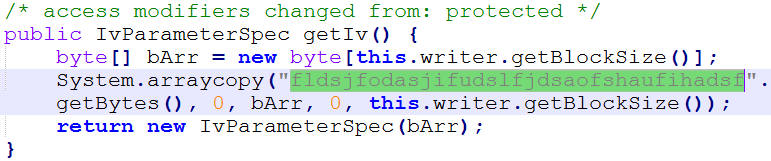}
  \caption{Use of hard-coded IVs.}
  \label{fig:insecure-hardcoded-iv}
\end{figure}

Among the prevalent types of threats, weak message confidentiality (ID\_df5) was verified in several apps due to the use of insecure cryptography, which leads to no channel confidentiality (ID\_df7).
Manual verification of the apps' reverse-engineered code was performed, revealing that fifteen apps used insecure PRNGs (e.g., see Figure \ref{fig:insecure-random-iv}).
Also, seven apps used insecure cyphers (i.e. MD5 and SHA1), and one app used an insecure cypher mode (ECB).
We also manually investigated insecure Initialisation Vectors (IVs) used in the apps. 
A total of 12 apps were found to have used insecure IVs (e.g., see Figure \ref{fig:insecure-hardcoded-iv}).

\begin{figure}[h]
  \centering
  \includegraphics[width=0.7\linewidth]{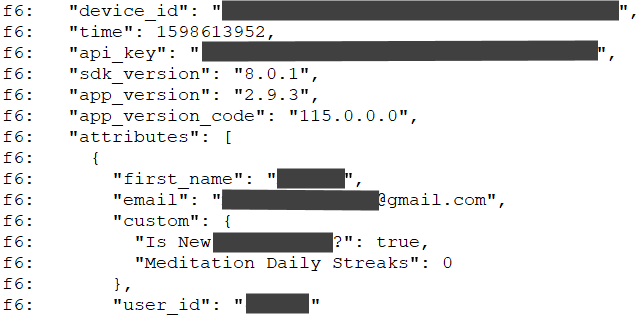}
  \caption{Example of sensitive information in Logcat logs.}
  \label{fig:logcat-sample}
\end{figure}

Another common threat is the lack of message confidentiality (ID\_df4). During the log analysis, we sought to identify four types of data leaks, as shown in Table \ref{tab:logcat-results}. These are alarming results as this information in Logcat logs can be accessed by other apps that are running in a device \citep{kotipalli2016hacking}. 
Figure \ref{fig:logcat-sample} shows an example of a Logcat log snippet identified to log personal data of the user and API keys.

\begin{table}[htbp]
\caption{Analysis of Logcat logs.}
\begin{center}
\begin{tabular}{cp{0.7\linewidth}}
\hline
\textbf{N. of Apps} & \textbf{Information disclosure issues} \\
\hline
19 & Revealing apps' behaviour, usage, and activities. \\
15 & Logging web traffic, parameters, Post value. \\
5 & Revealing personal information. \\
4 & Revealing tokens and credentials used by the app. \\
\hline
\end{tabular}
\label{tab:logcat-results}
\end{center}
\end{table}

Threats to the stored data were also common, e.g., bypass protection scheme (ID\_ds1), data intelligible (ID\_ds2), or un-encrypted (ID\_ds10). 
Only four apps have used encryption for storing files, and none have used encrypted databases. 
We found 15 apps that stored users' personal information (e.g. email, password, address) in files or databases. 
Such information can be accessible by unintended parties (e.g., in case of device theft or malicious backups). 

Disclosure of the credentials was also observed at various stages of the static and dynamic analyses.
This could lead to spoofing of an external entity (S) if an attacker can obtain legitimate credentials (S\_1) from an existing user (e.g., username and password) or service (e.g., API keys).
For instance, when inspecting the generated network traffic, we found that 13 apps reveal API keys used to access 3rd-party services, leading to unauthorized access to micro-services and APIs. 
Two apps also revealed the user email and password in the HTTP header or as GET parameters. 
Furthermore, 18 apps stored the credentials such as passwords, tokens and keys insecurely.

\subsubsection{Unawareness Threats}
\label{sec:unawareness-threats}
Unawareness refers to data subjects not being aware of the impacts and consequences of sharing personal data.
For instance, personal data is shared with mental health services and other services (i.e. cloud providers, analytics, advertising services).
In such cases, a system itself can support users in making privacy-aware decisions.
Such unawareness threats focus on a system's provisions to guide and educate users concerning their data sharing.

Evidence of unawareness threats was observed in the static and dynamic analyses, the requests of PIAs, and the communication with developers.
A type of threat concerns providing too much personal data (U\_1), which can be linked to the list of permissions required by the apps to run.
MobSF static analysis checks the apps for dangerous permissions, i.e., giving an app additional access to the restricted data and allowing an app to perform the restricted actions that substantially affect a system and other apps.
On average, the apps have 5.6 dangerous permissions ($std = 8.2$), with apps requiring a minimum of 3 up to 30 dangerous permissions.
Table \ref{tab:permissions} lists the most common dangerous permissions used by the studied apps.

\begin{table}[htbp]
\caption{Most common dangerous permissions used by apps.}
\begin{center}
\begin{tabular}{cp{0.5\linewidth}}
\hline
\textbf{N. of Apps} & \textbf{Dangerous Permissions} \\
\hline
27 & \texttt{android.permission.INTERNET} \\
24 & \texttt{android.permission.WAKE\_LOCK} \\
23 & \texttt{com.google.android.finsky.permission- .BIND\_GET\_INSTALL\_REFERRER\_SERVICE} \\
19 & \texttt{android.permission.WRITE\_EXTERNAL\_STORAGE} \\
16 & \texttt{com.android.vending.BILLING} \\
13 & \texttt{android.permission.READ\_EXTERNAL\_STORAGE} \\
9 & \texttt{android.permission.READ\_PHONE\_STATE} \\
7 & \texttt{android.permission.ACCESS\_FINE\_LOCATION} \\
6 & \texttt{android.permission.RECORD\_AUDIO} \\
6 & \texttt{android.permission.MODIFY\_AUDIO\_SETTINGS} \\
6 & \texttt{android.permission.CAMERA} \\
6 & \texttt{android.permission.ACCESS\_COARSE\_LOCATION} \\
\hline
\end{tabular}
\label{tab:permissions}
\end{center}
\end{table}

As mentioned in Section \ref{sec:methodology-static-analysis}, two authors manually inspected the dangerous permissions to verify whether they are necessary to serve the app's purpose.
During the evaluation, we used the apps in real mobile phones, and checked for functions that would justify the use of a given dangerous permission.
Dangerous permissions that did not seem necessary were flagged and included as a potential issue in the reports later sent to developers.
Most of the dangerous permissions were not deemed necessary for the apps to function.
For instance, the pair of permissions \texttt{READ\_EXTERNAL\_STORAGE} and \texttt{WRITE\_EXTERNAL\_STORAGE} are not always needed, but they are dangerous because they grant an app indiscriminate access to the device's external storage, where a user's sensitive information may be stored.
On average, the apps use 4.1 ($std = 7.6$) unnecessary dangerous permissions.
Even though software developers may have justifiable purposes for requiring such permissions, users must clearly understand them.

In this study, we also took the initiative of contacting the companies whose apps were studied and requesting the PIA reports of their respective apps.
This step revealed a degree of no/insufficient feedback and awareness tools (U\_3), considering that PIAs reflect on the impacts of information sharing.
Only three (11\%) companies carried out a PIA for their apps, and only two of them made the PIA report available to us.
Of the remaining companies, twenty (75\%) did not answer this PIA request, and four (15\%) reported not conducting a PIA.
It is worth mentioning that PIAs would help companies to demonstrate compliance to data protection authorities, which relates to the following subsection on Non-compliance threats.
Furthermore, if we consider mental health apps as likely to result in ``high-risk'' to the rights and freedoms of natural persons, PIAs are mandatory according to the EU GDPR \citep{wp292017pia}.

We can also consider the companies' feedback in the responsible disclosure process.
We emailed the evaluation reports consisting of all the issues found for different apps to their respective companies. We received responses from seven companies (26\%) that provided us with their feedback and the actions taken. The responses from software developers, lead engineers, and privacy officers were positive. They all showed appreciation to well-intended ethical researchers supporting them, with the desire to help build more secure and privacy-preserving apps. Three companies have reported back, saying that the raised issues were or are being fixed for the subsequent releases of the apps. One company also provided a detailed response, addressing all raised issues one by one.

\subsubsection{Non-compliance Threats}
\label{sec:non-compliance-threats}
Non-compliance refers to adherence to legislation, regulations, and corporate policies.
LINDDUN uses this threat category to cover privacy notices and policies that should be provided to all users to inform them about the data collected, stored, and processed by systems.
Privacy policies and consent are linked, given that users have to read and understand the apps' privacy policy to provide informed consent.

The analyses of the apps' privacy policies, using readability scores and AI-assisted privacy tools, allowed the identification of non-compliance threats concerning incorrect or insufficient privacy policy (NC\_2) and insufficient notice (NC\_4).
Considering the Flesch-Kincaid reading ease measurement, most apps (89\%, n = 27) scored between 30-50 in the readability index, meaning that their privacy policies are difficult to read, requiring college-level education. 
Three apps scored a 50-60 range index, implying that the privacy policies are reasonably challenging to read, requiring 10th- to 12th-grade level education.
Interestingly, only one app provided a layered privacy policy \citep{timpson2009layers}, providing a 1st-layer summary and a 2nd-layer with the complete privacy policy, making it easier to read and understand.

\begin{figure*}[h]
  \centering
  \includegraphics[width=1.0\linewidth]{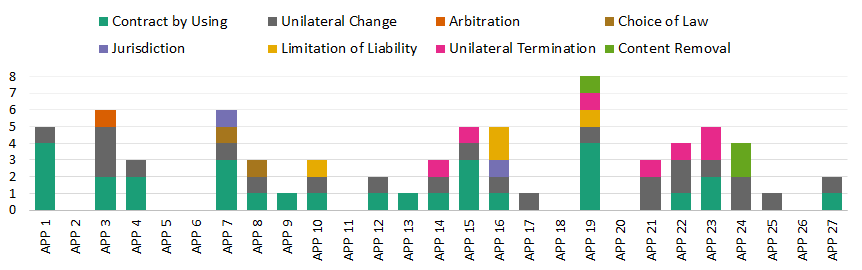}
  \caption{Summary of CLAUDETTE results of unfair clauses.}
  \label{fig:CLAUDETTE-result}
\end{figure*}

Threats in terms of incorrect or insufficient policies (NC\_2) were also revealed using the CLAUDETTE tool to identify unfair clauses.
Figure \ref{fig:CLAUDETTE-result} presents a summary of the results obtained using CLAUDETTE. 
On average, the apps' privacy policies had 2,7 unfair clauses
The most common type of unfair clause we observed was `Unilateral change', presented in the privacy policies of 18 apps. Furthermore, 16 privacy policies had unfair clauses in the `Contract by using' category.

We further analysed the apps' privacy policies using the PrivacyCheck tool, which scores the apps in terms of (1) user control over privacy and (2) GDPR compliance.
We used this tool to check the privacy policies of 26 apps, except for one app that the tool failed to interpret.
On average, the apps obtained a user control score of 59/100 ($std=15.14$), and a GDPR score of 63.1/100 ($std=31.25$).

\begin{figure}[h]
  \centering
  \includegraphics[width=0.6\linewidth]{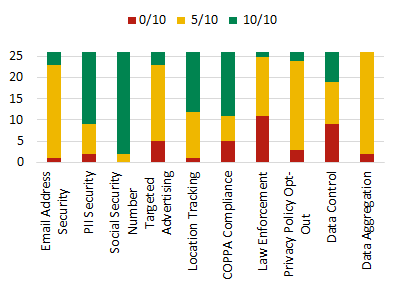}
  \caption{Summary of User Control scores from PrivacyCheck.}
  \label{fig:PrivacyCheck-control-result}
\end{figure}

Figure \ref{fig:PrivacyCheck-control-result} presents a more detailed summary of the PrivacyCheck scores obtained for the ten questions corresponding to the users' control. 
As shown in the figure, our sample of apps scored very poorly for questions such as \textit{``Does the site share your information with law enforcement?''} (11/26 apps scored 0/10) and \textit{``Does the site allow you to edit or delete your information from its records?''} (9/26 apps scored 0/10).
However, it appeared that the apps handled some privacy aspects more effectively, such as \textit{``How does the site handle your Social Security number?''} (24/26 apps scored 10/10) and \textit{``How does the site handle your credit card number and home address?''} (17/27 apps scored 10/10).

\begin{figure}[ht]
  \centering
  \includegraphics[width=0.6\linewidth]{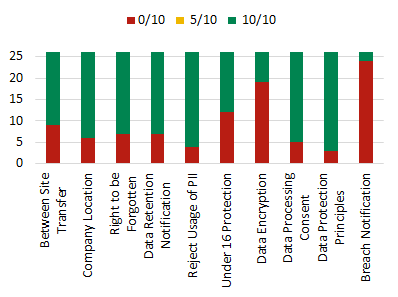}
  \caption{Summary of GDPR scores from PrivacyCheck.}
  \label{fig:PrivacyCheck-gdpr-result}
\end{figure}

Similarly, Figure \ref{fig:PrivacyCheck-gdpr-result} presents the PrivacyCheck scores obtained for the ten questions corresponding to GDPR compliance. 
The lowest compliance was observed for \textit{``Does the site notify the supervisory authority without undue delay if a breach of data happens?''} (24/26 apps scored 0/10) and \textit{``Does the site advise that their data is encrypted even while at rest?''} (19/26 apps scored 0/10). 
Most apps showed better compliance for questions such as \textit{``Does the site implement measures that meet the principles of data protection by design and by default?''} (23/26 apps scored 10/10) and \textit{``Does the site allow the user object to the use of their PII or limit the way that the information is utilized?''} (22/26 apps scored 10/10).

\section{Discussion}
\label{sec:discussion}
The study's results enable us to answer the research question: \textit{What is the current privacy status of top-ranked mental health apps?}
Tables \ref{tab:main-findings} and \ref{tab:main-findings-continuation} summarise the most common privacy issues and their prevalence in the studied mental health apps, contextualising findings according to the LINDDUN threat categories.
Based on that, this section discusses the following concerning topics: (1) privacy impacts of mental health apps; (2) apps' permissions and data access; (3) apps' security testing and coding; (4) Privacy Impact Assessments; (5) privacy policies; and, (6) recommendations.

\begin{table*}[htbp]
\caption{Summary of main findings according to LINDDUN.}
\begin{center}
\begin{tabular}{p{0.425\linewidth}p{0.325\linewidth}c}
\hline
\textbf{Main Findings}  & \textbf{Threat Examples} & \textbf{LINDDUN}\\
\hline \hline
\textbf{Finding 01:} Out of the 27 top-ranked mental health apps selected, most of them address the conditions of anxiety, stress, burnout and depression. Also, over a third of them address various other mental disorders, e.g., addictions, bipolar, self-harm, PTSD and OCD. Hence, these apps' processing operations result in ``high-risk'' to the rights and freedoms of natural persons. & \textit{(finding is too general to be mapped)} & $\square \square \square \square \square \square \square $ \\
\textbf{Finding 02:} 74\% ($n = 20$) of the apps scored as Critical Risk and 15\% ($n = 4$) as High Risk in the App Security Score during the static analysis. & \textit{(finding is too general to be mapped)} & $\square \square \square \square \square \square \square $ \\ 
\textbf{Finding 03:} All apps require dangerous permissions to run. Our manual inspection points to an average of 4 unnecessary dangerous permissions used being used, in which read/write operations to the external storage are of primary concern. & Providing too much personal data (U\_1), incorrect or insufficient privacy policies (NC\_2), insufficient notice (NC\_4). & $\square \square \square \square \square \blacksquare \blacksquare $ \\ 
\textbf{Finding 04:} Manual verification of the apps’ codes shows a high prevalence of fundamental secure coding problems related to the use of insecure PRNGs (56\%), insecure cyphers(56\%), insecure cypher modes (26\%), and insecure IVs (44\%). & Weak message confidentiality (ID\_df5). & $\square \square \square \square \blacksquare \square \square $ \\ 
\textbf{Finding 05:} 96\% ($n = 26$) of the apps contained hard-coded sensitive information like usernames, passwords and keys. & Spoofing an external entity (S), obtain legitimate credentials (S\_1). & $\square \square \square \square \blacksquare \square \square $ \\ 
\textbf{Finding 06:} Two apps reveal user email and credentials in the HTTP header or as GET parameters. & Spoofing an external entity (S), no message confidentiality (ID\_df4), no channel confidentiality (ID\_df7). & $\square \square \square \square \blacksquare \square \square $ \\ 
\textbf{Finding 07:} Two apps made a perplexing number of HTTP(S) requests in a short period. & Linkability and identifiability of contextual data (L\_df2, I\_df2), disclosure of a decrypted log of network connections (NR\_df7), timing requests visible (D\_df13), unaware of stored data (U\_2), insufficient notice (NC\_4). & $\blacksquare \blacksquare \blacksquare \blacksquare \square \blacksquare \blacksquare $ \\ 
\textbf{Finding 08:} User behaviour can also be easily extracted from web traffic logs (i.e., it is easy to perform profiling of mental health  apps’ users). & Linkability and identifiability of contextual data (L\_df2, I\_df2), disclosure of a decrypted log of network connections (NR\_df7), timing requests visible  (D\_df13), unaware of stored data (U\_2), insufficient notice (NC\_4). & $\blacksquare \blacksquare \blacksquare \blacksquare \square \blacksquare \blacksquare $ \\ 
\textbf{Finding 09:} 68\% ($n = 13$) of the apps reveal API keys used to access 3rd-party services. & Spoofing an external entity (S), obtain legitimate credentials (S\_1). & $\square \square \square \square \blacksquare \square \square $ \\ 
\hline
\end{tabular}
\label{tab:main-findings}
\end{center}
\end{table*}

\begin{table*}[htbp]
\caption{(Continuation) Summary of main findings according to LINDDUN.}
\begin{center}
\begin{tabular}{p{0.425\linewidth}p{0.325\linewidth}c}
\hline
\textbf{Main Findings}  & \textbf{Threat Examples} & \textbf{LINDDUN}\\
\hline \hline
\textbf{Finding 10:} Most apps try to pseudo-anonymize users through and anonymous IDs or hashed advertisement IDs, but these IDs can still be used to link data among various 3rd-parties. & Linkability and identifiability of contextual data (L\_df2, I\_df2), person wanting deniability cannot edit database (NR\_ds3), timing requests visible (D\_df13), unaware of stored data (U\_2), insufficient notice (NC\_4). & $\blacksquare \blacksquare \blacksquare \blacksquare \square \blacksquare \blacksquare $ \\ 
\textbf{Finding 11:} Apps communicate with a large number of 3rd-parties, for marketing and advertising, cloud service pro-visioning, and payments services. & Person wanting deniability cannot edit database (NR\_ds3), unaware of stored data (U\_2), insufficient notice (NC\_4). & $\square \square \blacksquare \square \square \blacksquare \blacksquare $ \\ 
\textbf{Finding 12:} All analyzed apps reveal users’ usage and apps’ behaviour in the Android system logs (i.e., Logcat), which is visible to all applications in the system. & Weak access control to data (base) (I\_ds1), bypass protection scheme (ID\_ds1), data intelligible (ID\_ds2). & $\square \blacksquare \square \square \blacksquare \square \square $ \\ 
\textbf{Finding 13:} 79\% ($n = 15$) of the apps store data in plain-text in the file system or in databases. & Weak access control to data (base) (I\_ds1), unencrypted (ID\_ds10). & $\square \blacksquare \square \square \blacksquare \square \square $ \\ 
\textbf{Finding 14:} 95\% ($n = 18$) of the apps reveal some credentials (e.g., API keys and tokens) in the stored data. & Weak access control to data (base) (I\_ds1), obtain legitimate credentials (S\_1). & $\square \blacksquare \square \square \blacksquare \square \square $ \\ 
\textbf{Finding 15:} 79\% ($n = 15$) of the apps’ databases are not encrypted. & Weak access control to data(base) (I\_ds1), unencrypted (ID\_ds10). & $\square \blacksquare \square \square \blacksquare \square \square $ \\ 
\textbf{Finding 16:} Twenty apps (75\%) did not report whether they conducted or not a PIA, and 4 (15\%) apps explicitly declared not conducting a PIA. & No/insufficient feedback and awareness tools (U\_3), insufficient notice (NC\_4). & $\square \square \square \square \square \blacksquare \blacksquare $ \\ 
\textbf{Finding 17:} Flesch-Kincaid Reading Ease average is 42 (i.e., Difficult to read) for the cohesiveness and complexity of the apps’ privacy policies. & Incorrect or insufficient privacy policy (NC\_2), insufficient notice (NC\_4). & $\square \square \square \square \square \square \blacksquare $ \\ 
\textbf{Finding 18:} An average of 2,7 unfair clauses was revealed for the  analysed privacy policies. & Incorrect or insufficient privacy policy (NC\_2), insufficient notice (NC\_4). & $\square \square \square \square \square \square \blacksquare $ \\ 
\textbf{Finding 19:} The average user control score that the apps obtained was 59/100 ($std = 15.14$), and the average GDPR score was 63.1/100 ($std = 31.25$). & Incorrect or insufficient privacy policy (NC\_2), insufficient notice (NC\_4). & $\square \square \square \square \square \square \blacksquare $ \\ 
\textbf{Finding 20:} 74\% ($n = 20$) of the companies have not replied to the reports sent for responsible disclosure. & No/insufficient feedback and awareness tools (U\_3), insufficient notice (NC\_4). & $\square \square \square \square \square \blacksquare \blacksquare $ \\ 
\hline
\end{tabular}
\label{tab:main-findings-continuation}
\end{center}
\end{table*}

\subsection{Privacy Impacts of Mental Health Apps}
Even though mental health apps have higher privacy impacts, the results show that these apps contain most of the privacy and security issues found in an average Android app.
For example, our analysis identified vulnerabilities related to all seven Android app vulnerability categories (i.e., cryptography API, inter-component communication, networking, permission, data storage, system processes, and web API) presented by \cite{ranganath2020free}.
Furthermore, various privacy issues were identified, such as insufficient levels of information handling, similar to what other researchers have observed in different types of mobile apps \citep{huckvale2019assessment, powell2018complexity}.

Privacy violations in mental health apps tend to have severe negative impacts on the rights and freedoms of natural persons, therefore calling for higher levels of protection and safeguards. 
Some issues identified in this privacy analysis would have a lower impact in a general Android app (e.g., WhatsApp, Twitter, Netflix apps). 
For example, disclosure of identifiers to 3rd-parties, such as IMEI, UUID and IP address, would have a low impact in a general app. 
Perhaps, most users would not even consider it as an issue.
In contrast, mental health app users would consider this invasive since most users would not even want other people to know that they are using mental health apps. 
Research has shown that breaches of mental health information have severe repercussions, such as exploitative targeted advertising and negative impacts on an individual's employability, credit rating, or ability to access rental housing \citep{parker2019private}.

\subsection{Apps' Permissions and Data Access}
\label{sec:discussion-permissions}
During the static analysis, we found that all apps use one or more dangerous permissions.
Many of these permissions could be avoided or at least better explained to end-users.
For instance, the pair of dangerous permissions \texttt{READ\_EXTERNAL\_STORAGE} and \texttt{WRITE\_EXTERNAL\_STORAGE}.
Based on our manual analysis of apps' permissions (Section \ref{sec:unawareness-threats}), we noticed that the apps rarely need access to external storage.
Thus, these permissions could have been avoided or more carefully used.

The apps also request such permissions (i.e., get user approval) when they are first opened.
Users can indeed revoke dangerous permissions from any app at any time (i.e., if they know how to do it).
However, it would be recommended that app developers ask for permissions ``in context'', i.e., when the user starts to interact with the feature that requires it.
Also, if permissions are not essential for the apps to function, they could be disabled by default, i.e., running the app most privately.

Future research could also focus on the apps' permissions, data access, and sharing behaviours over more extended periods.
For instance, similar to \citep{momen2020measuring}, in which researchers
had apps installed on real devices over time (e.g., months) analysing the apps' behaviour under various conditions.
Ideally, developers would benefit the most if they could rely on a testbed for privacy assessment, as the one proposed by the REsearch centre on Privacy, Harm Reduction and Adversarial INfluence online (REPHRAIN) \citep{gardiner2021building}.
Such testbed would enable developers to only drop their app file into an user interface, following a wizard-based tool. 
The testbed then runs multiple static and dynamic privacy tests against the file and produces a report in a comma-separated format, which the developer can download for their own analysis.

\subsection{Apps' Security Testing \& Coding}
The results from our static analysis (Section \ref{sec:disclosure-info-threats}) showed that most apps are at critical risk ($n=20$) or high risk ($n=4$).
Vulnerabilities such as hard-coded secrets \citep{lee2019identifying}, use of weak algorithms and protocols (ECB, TLS 1.0, etc.), weak IVs, and insecure PRNGs \citep{egele2013empirical} were also verified. 
The MobSF tool has been continuously upgraded, making such security testing relatively straightforward.
App developers could have identified most of these issues using MobSF's automated static analysis.
The prevalence of such vulnerabilities suggests that app developers are not adhering to the basic principles of secure coding.
Furthermore, it is worth stressing that many of our findings were identified in the dynamic analysis.
The inspection of network traffic, stored data, and logs can reveal several issues that a static analysis alone cannot.

A recent study found that 85\%  of mHealth developers reported little to no budget for security \citep{aljedaani2020empirical} and that 80\% of the mHealth developers reported having insufficient security knowledge \citep{aljedaani2020empirical, aljedaani2021challenges}.
We believe that the developers of the mHealth apps analyzed in this study faced similar challenges that are also evident from the following observations concerning secure coding/programming.
First, the use of insecure randoms, cypher modes, and IVs, i.e., incorrect use of cryptographic components.
Second, the insecure logs, leaking the app's behaviour and the user's data, either internally to the system logs (e.g., Logcat) or externally to cloud-based logging services (e.g. Loggly).
Third, the presence of hard-coded information, such as tokens and API keys.
Such findings signal that app developers require more security training and that security testing may not be part of the development process.

\subsection{PIAs \& DPIAs}
From the start, we contacted all the relevant companies whose apps we selected for study for obtaining the PIAs reports (if available). However, we received only two public PIA reports.
These PIA reports were relatively brief, lacking sufficient information about the apps' systems and components.
PIAs should usually start with a high-level data flow diagram that shows what personal data is collected and how it is processed and shared among 3rd-party services \citep{wp292017pia}.
We assert that it is important for an mHealth app to identify the potential privacy threats and apply suitable countermeasures for eliminating or mitigating the identified risks during appropriate phases of development/evolution.
As per our findings, a large majority of the MHealth apps developers seem to be unaware of the PIA requirements that are usually mandatory according to some regulations, such as GDPR.

Whilst it is understandable that performing and updating full-fledged PIAs is a time-consuming process , e.g., see the PIA \citep{iwaya2019pia}, mHealth apps development companies and developers can benefit from the available knowledge resources and guidelines such as the work reported in this report \citep{mantovani2017code}.
The knowledge and time invested in performing PIA and making that public will help increase the trust of the end users and the relevant authorities.

\subsection{Privacy Policies: Transparency, Consent and Intervenability}
All the analysed mHealth apps had a privacy policy.
This is quite positive if compared to other studies that reported that only 46\% of dementia apps \citep{rosenfeld2017dementia} and 19\% of diabetes apps \citep{blenner2016privacy} had a privacy policy.
This is likely because we analyzed only top-ranked apps with a large user bases.
However, the readability scores of the privacy policies are still low.
According to other studies, the average grade-level readability should be calculated as the average of the scores from the Gunning Fog, Flesh-Kincaid Grade Level, and SMOG formulas \citep{robillard2019availability, sunyaev2014availability}.
In such case, the average grade-level readability for the analyzed privacy policies was 13.21, consistent with the scores of 13.78 in \citep{robillard2019availability} and 16.00 in \citep{sunyaev2014availability}.
Privacy policies are still hard to read, raising concerns with regards to transparency and consent.

Privacy policies also present unfair clauses, of which ``contract by using'' and ``unilateral change'' are the two most common types.
Contract by using is incredibly unfair in the case of mHealth apps.
Such apps should rely on explicit informed consent since they handle sensitive personal data of people who may be considered relative more vulnerable and fragile.
The EU GDPR (Art. 4 (11) defines consent as freely given, specific, informed and with explicit indication of the data subject's wishes to use the system and have his or her data collected and processed \citep{GDPR2016}.
Contract by using defies this idea of consent.
Companies should review their apps' privacy policy and, most importantly, change the apps to honestly inform users, recording their consent to collect and process data.

Most apps' consent process was just an initial screen presenting the privacy policy and an ``I agree'' button.
Understandably, developers design their apps with as few steps as possible in the onboarding process, reducing friction and improving users' experience.
However, poor privacy also causes a bad user experience.
Balancing privacy and user experience is challenging and demands further investigation.
However, developers could ask themselves: \textit{``Would my users be surprised if they knew about all the data that is collected, the processing purposes, or the extent of data sharing?''}
Any privacy ``surprises'' reveal issues that need to be raised and discussed, users should be informed, and the system's design should be reviewed.

For instance, many mHealth apps rely on advertising as monetary revenue.
Users of mHealth apps, even if de-identified, are still targeted with personalised advertisements based on their unique ``anonymous'' IDs (e.g., \texttt{uuid} and \texttt{aaid}).
Also, the advanced paradigms of personal advertising, such as cross-device tracking (CDT), are commonly used to monitor users' browsing on multiple devices and screens for delivering (re-)targeted ads on the most appropriate screen.
For instance, if a person downloads an mHealth app on one's mobile device, it is likely that person will see other ads about mental health in one's Facebook timeline when using a PC.
Researchers have already found that CDT undoubtedly infringes users' online privacy and minimizes their anonymity \citep{solomos2019talon}.
Besides, there is a risk of exploitative advertising to individuals who may be vulnerable due to mental health conditions.
Such extent of data processing is likely to surprise users (and developers), unaware of privacy risks and impacts.
These observations enable us to support the growing arguments that apps development is intrinsically linked to the online advertising businesses, which may give little to no control on the management and utilization of data to those from whom the data is gathered, i.e., end users.

\section{Limitations}
\label{sec:limitations}
Some limitations in terms of the methodology need to be considered when interpreting the results and findings of this study.
We manually investigated the code snippets flagged for insecure PRNGs, cyphers, and cypher modes during the static analysis. 
That is, we limited our analysis to the files flagged by MobSF.
However, we observed that some of the reported code snippets used insecure PRNGs and cyphers to create \texttt{wrapper} classes and \texttt{util} methods for the original functionality. 
Even though using these \texttt{wrapper} classes and \texttt{util} methods in security contexts would lead to a security vulnerability, our analysis did not investigate such usages as it would increase the complexity and time required for the study. 
We have shared this observation with the studied apps' development companies as part of the responsible disclosure process and advised them to consider it when interpreting the reported results.

During the dynamic analysis, some apps were not compatible to run on the Genymotion emulator with MobSF.
Hence, the results are limited to a smaller sample of 19 apps that were fully dynamically analysed.
This process required the manual operation of the apps, attempting to cover all of the accessible functionalities. However, we neither performed any credit card payments nor paid to test the premium features, limiting the extent of testing.

Regarding the analysis of the privacy policies, we relied on two AI-based tools: (1) CLAUDETTE, to identify unfair clauses; and, PrivacyCheck, to calculate user's control and GDPR compliance scores.
Although such tools give us a metric for comparison, an ideal analysis of privacy policies would require legal analysis of the text made by a privacy lawyer.
These AI-based tools also have some limitations concerning their accuracy.
According to the creators of these tools, CLAUDETTE has an accuracy of 78\% for identifying unfair clauses and an accuracy between 74\%-95\% for distinguishing between unfair clause categories \citep{lippi2019claudette}.
PrivacyCheck has an accuracy of 60\% when scoring privacy policies for the ten user control questions and the ten GDPR questions \citep{zaeem2020privacycheck}.
Thus, results should be interpreted with such limitations in mind.

\section{Conclusion}
\label{sec:conclusion}
Mental health apps offer new pathways for people to seek psychological support anywhere and anytime. The innovative use of technological advances in mobile devices for providing mental health (or well-being) support purports to significantly improve people's quality of life. However, the mobile apps are increasingly vulnerable to data privacy breaches as a result of security attacks. A data privacy breach of an app may result in financial, social, physical or mental stress. Given the users of mental health apps are usually facing psychological issues such as depression, anxiety and stress, the detrimental impact of an app's data privacy breach can have more significant negative impact on users. 
Thus, it is of utmost importance that the development of mHealth apps follow the practices that ensure privacy by design. 

We decided to empirically study the data privacy of mental health apps. Our empirical investigation shows a high prevalence of information disclosure threats, mainly originating from insecure programming.
Threats related to linkability, identifiability, non-repudiation and detectability are also exacerbated by the large number of third parties in the apps' ecosystem, facilitating profiling of users and exploitative advertising.
Apps also lack transparency and sufficient notice mechanisms, leading to unawareness and non-compliance threats.

This study has provided us with sufficient empirical evidence to assert that mobile apps in general but mental health apps in particular ought to be developed by following privacy by design paradigm. Moreover, this study has also enabled us to surmise that apart from developers, other stakeholders can also play important roles in ensuring data privacy in mHealth apps. Based on this research, we have compiled a list of data-informed actionable measures as a set of recommendations for ensuring data privacy in mHealth apps. Tables~\ref{tab:recommendations} and \ref{tab:recommendations-continuation} provide the list of recommendations linked to the findings presented in Table~\ref{tab:main-findings}. 
We expect that these recommendations will enable all the key stakeholders, particularly the apps developers, to play their respective parts in order to ensure the privacy of the data of mHealth apps.

\begin{table}
\caption{Summary of recommendations to multiple stakeholders.}
\begin{center}
\begin{tabular}{p{0.85\linewidth}p{0.075\linewidth}}
\hline
\textbf{Organizations} & \textbf{Find.}\\
\hline
$\longrightarrow$ \textbf{Undertake a Privacy Impact Assessment} -- Demonstrate compliance by conducting Privacy Impact Assessment, even if not a full-fledged PIA. There are more concise/simplified methodologies for mHealth. & 16 \\
$\longrightarrow$ \textbf{Assume Mental Health Apps as High-Risk Systems} -- Development processes should be fine-tuned to give better emphasis to security and privacy. When developing a health app (or mental health app), higher levels of security and privacy should be considered compared to other general apps. & 1, 8 \\
$\longrightarrow$ \textbf{Engage with Experts} -- Better engagement of security and privacy experts in the development and evaluation, as well as in the writing of privacy policies to avoid unfair clauses. & 20 \\
$\longrightarrow$ \textbf{Write Readable Privacy Policies} -- Enhance transparency and openness by writing accessible Privacy Policies that truly allow users to understand and make informed decisions. & 17 \\
\hline \hline
\textbf{Software developers} & \textbf{Find.}\\
\hline
$\longrightarrow$ \textbf{Beware of the Unskilled and Unaware} -- Likely, app developers do not know the extent of security and privacy risks of using 3rd-party SDKs and APIs. That, matched with the lack of security knowledge, might make them prone to a Dunning-Kruger effect on security knowledge, i.e., overseeing and underestimating security and privacy issues while also overestimating their levels of secure coding abilities \citep{ament2017ubiquitous,wagner2019tooconfident}. & 4, 5, 6, 9, 12, 13, 14, 15 \\
$\longrightarrow$ \textbf{Connect the Privacy Policy to the System's Design} -- Even though privacy policies are not within the software developers responsibility, they should be familiar with their app's privacy policy and terms of service. Interact with lawyers (or whoever is responsible for writing and updating the privacy policy) whenever necessary to correct information on data collection, purpose limitation and specification, and ensure security and privacy by design. & 17, 18, 19\\
$\longrightarrow$ \textbf{Engineer Privacy By Design and By Default} (Art. 25 GDPR \citep{GDPR2016}) -- Software developers should be aware that the GDPR states that \textit{``controller shall implement appropriate technical and organisational measures''}. Even though implementing the ``state-of-the-art'' is not always ``technically'' possible in all organisations and systems, vulnerabilities related to very basic secure coding practices are rather concerning. & 2, 4, 7, 9, 13, 14, 15 \\
$\longrightarrow$ \textbf{Collect Valid Consent with Responsible On-boarding} -- Even though the use of proper consent mechanisms may add friction to the on-boarding process, mental health apps rely on user consent to operate, so it is important that valid consent is being collected. After consent, apps should operate in the most privacy-preserving way by default (e.g., no advertising), and the consent withdrawal should be as easy as providing consent. & 8, 10, 11, 18 \\
\hline \hline
\textbf{End-users and Health Practitioners} & \textbf{Find.}\\
\hline
$\longrightarrow$ \textbf{Stand Up for Your Rights} -- Users that value their privacy can exercise rights by requesting more privacy-friendly apps. Users can question the current privacy policies and consent mechanisms. Request access to their data and better information on the nature of the data processing. & 10, 11, 16, 19 \\
$\longrightarrow$ \textbf{Recommend Reputable Apps for Mental Health} -- Health practitioners should encourage their patients to take higher control over their treatment and journey towards better mental health. Mental health apps can help with that, but practitioners should pay careful attention and recommend only apps that respect users' privacy. & 16, 19 \\
\hline
\end{tabular}
\label{tab:recommendations}
\end{center}
\end{table}

\begin{table}
\caption{(Continuation) Summary of recommendations to multiple stakeholders.}
\begin{center}
\begin{tabular}{p{0.85\linewidth}p{0.075\linewidth}}
\hline
\textbf{Mobile App Distribution Platforms} & \textbf{Find.}\\
\hline
$\longrightarrow$ \textbf{Raise the Bar for High-Risk Apps} -- App distributors could require better privacy measurements to be put in place. Distributors could also categorise high-risk apps, adding filters for health genre apps. & 1, 16 \\
$\longrightarrow$ \textbf{Enhance Trust and Transparency} \citep{bal2014user} -- App distributors could also add useful privacy information about apps, especially about privacy consequences to support decision-making, and add privacy ratings for apps based on their data-access profiles and purposes of data access. & 8, 10, 11, 12 \\
\hline \hline
\textbf{Smartphone Platform Providers} & \textbf{Find.}\\
\hline
$\longrightarrow$ \textbf{Call for Privacy-Friendly System Apps and API Frameworks} \citep{bal2014user} -- Smartphone providers could develop common systems to keep track of sensitive information flows, as well as to communicate observed behaviour to users, and provide developers with standardised ways to explain permission requests. & 3 \\
\hline
\end{tabular}
\label{tab:recommendations-continuation}
\end{center}
\end{table}

These recommendations also serve to reiterate the fact that developers alone cannot implement all the safeguards to mitigate, reduce or eliminate the identified threats.
The companies' leaders and top management are the ones who define the business models around the mental health apps.
For instance, when considering the excessive use of 3rd-parties and data brokers, the software developers might be able to raise privacy issues, but it is ultimately the responsibility of the leaders to re-think and adopt more privacy-preserving business strategies.
Simply put, no amount of technical and organisational privacy controls can fix a broken business model that inherently undermines people's privacy.

This empirical study suggests that companies and app developers still need to be more knowledgeable and experienced when considering and addressing privacy risks in the app development process.
At the same time, leaders and managers need to review their business models and re-think their design practices in the organisations.
Raising awareness among users and health professionals is also crucial.
Users should drive the demand for more privacy-preserving apps.
Mental health professionals should carefully evaluate the apps to recommend privacy-friendly and safe apps to their clients

Besides, there are also initiatives that the app distribution platforms (e.g., Google Play Store) and the smartphone platform providers (e.g., Android) could take to enhance privacy in the ecosystem.
App stores could increase the vetting process for high-risk apps, such as those in medical, health and fitness application categories.
Also, as suggested by \cite{bal2014user}, the app stores could provide more helpful privacy information about the apps (e.g., using privacy rating scale), and smartphone platforms could provide privacy-enhancing mechanisms in the operational systems.

\begin{acknowledgements}
The work has been supported by the Cyber Security Cooperative Research Centre (CSCRC) Limited, whose activities are partially funded by the Australian Government’s Cooperative Research Centres Programme.
We also thank Dr Minhui Xue (Jason Xue) and Dr Constantinos Patsakis for their early advice regarding the implications of performing security testing of mHealth apps on the market.
\end{acknowledgements}

\bibliographystyle{spbasic}      % basic style, author-year citations
\bibliography{myrefs}   % name your BibTeX data base

\end{document}